\documentclass[10pt,aps,prd,onecolumn,superscriptaddress,showpacs,showkeys,nofootinbib,floatfix,preprintnumbers]{revtex4}
\usepackage{appendix}
\usepackage{epsfig}
\usepackage{graphicx}
\usepackage{amsfonts}
\usepackage{amsmath}
\usepackage{psfrag}
\usepackage{amssymb,amsmath,amsthm}
\usepackage{verbatim}
\usepackage{soul}
\usepackage[normalem]{ulem}
\usepackage{color}
\usepackage{multirow}
\usepackage{titlesec}

\begin{document}

\title{Nonequilibrium pion distribution within the Zubarev approach}

% % Author Orchid ID: enter ID or remove command
% %\newcommand{\orcidauthorA}{0000-0002-5380-0728} % Add \orcidA{} behind the author's name
% \newcommand{\orcidauthorA}{0000-0002-8399-5183} % Add \orcidA{} behind the author's name
% %\newcommand{\orcidauthorC}{0000-0000-000-000X} % Add \orcidB{} behind the author's name
% %\newcommand{\orcidauthorD}{0000-0000-0000-0000} % Add \orcidD{} behind the author's name

% % Authors, for the paper (add full first names)
\author{David Blaschke}
\affiliation{Institute of Theoretical Physics, University of Wroclaw, PL-50204 Wroclaw, Poland}
\affiliation{Bogoliubov Laboratory of Theoretical Physics, Joint Institute for Nuclear Research, 141980 Dubna, Russia}
\affiliation{National Research Nuclear University (MEPhI), 115409 Moscow, Russia}
\author{Gerd R\"opke} 
\affiliation{National Research Nuclear University (MEPhI), 115409 Moscow, Russia}
\affiliation{Institute of Physics, University of Rostock, 18059 Rostock, Germany}
\author{Dmitry N. Voskresensky}
\affiliation{Bogoliubov Laboratory of Theoretical Physics, Joint Institute for Nuclear Research, 141980 Dubna, Russia}
\affiliation{National Research Nuclear University (MEPhI), 115409 Moscow, Russia}
\author{Vladimir G. Morozov}
\affiliation{MIREA-Russian Technological University, %Vernadsky Av. 78, 
119454 Moscow, Russia}

% 

% Contact information of the corresponding author
%\corres{Correspondence: david.blaschke@gmail.com}

\begin{abstract}
We discuss how the nonequilibrium process of pion production within the Zubarev approach of the nonequilibrium statistical operator leads to a theoretical foundation for the appearance of a nonequilibrium pion chemical potential for the pion distribution function for which there is experimental evidence in experiments at the CERN LHC.
\end{abstract}
% Keywords
% The fields PACS, MSC, and JEL may be left empty or commented out if not applicable
\maketitle

%%%%%%%%%%%%%%%%%%%%%%%%%%%%%%%%

\section{Introduction}

The thermal statistical model 
\cite{BraunMunzinger:2001ip,Andronic:2017pug,Stachel:2013zma,Becattini:2012xb,Cleymans:2006xj,Cleymans:2010qya}
for chemical freeze-out of hadron species gives a successful description of a
set of particle ratios produced in heavy-ion collisions (HIC) at different center of mass energies ranging from 
SIS-18 over AGS and SPS energies up to RHIC and LHC. 
It came therefore as a surprise that for LHC at $\sqrt{s}=2.76$ TeV the measured proton abundances 
\cite{Abelev:2012wca,Abelev:2013vea} do not agree with the most common version of the thermal model\footnote{The inclusion of resonance formation due to (multi-)pion-nucleon interaction and further correlations in the continuum
within the Beth-Uhlenbeck approach \cite{Weinhold:1997ig,Andronic:2018qqt} improves the agreement with the experiment.} 
based on the grand canonical ensemble \cite{Stachel:2013zma,Becattini:2012xb}.
As a possible explanation of this effect it has been suggested that the freeze-out may take place off chemical equilibrium \cite{Petran:2013qla,Petran:2013lja,Begun:2013nga}. 
Hereby, a key feature is the enhancement of low-transverse momentum pion spectra above the expectation 
from equilibrium statistical models which was seen already at lower energies in pion spectra at SPS and 
clearly seen in the RHIC and the LHC data.
The effect can be seen as a precursor of pion Bose-Einstein condensation due to high phase space occupation at low momenta and has consequently been parametrized by adopting a pion chemical potential very close to the pion mass 
\cite{Kataja:1990tp,Gavin:1991ki}.
This concept is based on the assumption  that the total pion number is dynamically fixed on a time scale between the pion chemical freeze out $t_{\pi,{\rm cfo}}$ and the thermal freeze-out (or simply freeze-out) $t_{\rm fo}$, 
$t_{\pi,{\rm cfo}}<t<t_{\rm fo}$, where at $t_{\pi,{\rm cfo}}$ the  pion number becomes frozen and at $t_{\rm fo}$ the momentum distributions stop to change~\cite{Gerber:1990yb}. 
Thereby, for pions we assume dominance of elastic collisions over inelastic ones. 
We will assume that for pions the time typical for absorptive processes $t_{\pi,{\rm abs}}$, which change the pion particle number, is $t_{\pi,{\rm abs}}>t_{\rm fo}$. 

%In the present paper we want to elucidate the relationship between the nonequilibrium processes taking place in the hadronizing QGP fireball 
%and the emergence of a nonequilibrium chemical potential for hadrons, in particular for the pions. 
In the present paper we want to elucidate the nonequilibrium evolution of the initial fireball 
and the emergence of a nonequilibrium chemical potential for hadrons, in particular for the pions. 
The theoretical background shall be provided by the Zubarev formalism of the nonequilibrium statistical operator (NSO)
\cite{Zubarev:1996} 
which introduces a generalization of the thermodynamical Gibbs ensemble by including nonequilibrium observables into 
the derivation of the statistical operator of the nonequilibrium, see also \cite{Zubarev:1971}.
This is facilitated by extending the set of Lagrangian multipliers by the additional nonequilibrium chemical potentials for the hadrons will appear in the NSO. 
If the nonequilibrium chemical potential for the pions coincides with the pion effective mass, Bose-Einstein condensation 
will occur, and strong effects are expected on the measured pion spectra \cite{Voskresensky:1994,Kolomeitsev:2017wjd}.
%when the nonequilibrium pion chemical potential would equal the pion mass.
%we will focus in particular on the case of the pion chemical potential 
%which is expected and actually exhibits the strongest effects also on the spectra due to the proximity of the Bose pole for pions 
%when the nonequilibrium pion chemical potential would equal the pion mass.

Many formulations have been given for the approach to equilibrium \cite{Welke:1990za,Fore:2019wib}
and the kinetics of Bose-Einstein condensation, see for instance 
Refs.~\cite{Semikoz:1994zp,Semikoz:1995rd,Voskresensky:1996ur,Berges:2018cny,Berges:2017hne}.
New ans\"atze have been developed, e.g.,  in Ref.~\cite{Erne:2018gmz,Mazeliauskas:2018yef} with a state that is a 
fixed point and the evolution towards it is universal. 
Via this fixed point the system develops then dynamically. 
 
Such a behavior is long known in the context of the Zubarev formalism, which is able to describe, for example, the transition from the kinetic stage to the hydrodynamic stage. 
The short-time evolution goes to a relevant statistical operator with local time-dependent thermodynamic parameters.
The long-time scale evolution is given by the time dependence of these  thermodynamic parameters approaching thermodynamic equilibrium. 
% Part of the dynamics is described microscopically (strong interaction) other part coupling to other particles (decay to electroweak) leads to dissipation, can be formulated by quantum master equation (dissipator). Here we make it differently using the Zubarev formalism, coupling to a bath. 

% 
% Appearance of a condensate in the entropy. Better than mean-field, GG or GG0.\\
% 
% Relaxation to relevant state. Pion as entropy (better gluon).\\
% 
% Question: role of Delta resonances? Bound state N+Pi, decay to N? Conservation of particle number? 
% PDG: decay of Delta to N+Pi 99.5 \%, to N+ gamma 0.5 \%. [Bijker et al., Revista Mexicana de F?sica 42,  Suplemento 1 (1996) 9-20].
% Pi+Pi: elastic or inelastic?

%%%%%%%%%%%%%%%%%%%%%%%%%%%%%%%%%%%%%%%%%%%%%%%%%%%%
\section{The nonequilibrium statistical operator method}
\label{sec:noneq}

%\subsection{Nonequilirium statistical approach}
%%%%%%%%%%%%%%%%%%%%%%%%%%%%%%%%%%%%%%%%%%%%%%%%%%%%
HIC at ultrarelativistic energies are violent nonequilibrium processes which need, in principle, a genuine nonequilibrium approach.
At present, simple approximations are used such as transport models based on the kinetic equations 
for single-particle distribution functions.
Transport codes based on the relativistic BUU or VUU equations have been worked out \cite{Welke,Semikoz:1994zp,Semikoz:1995rd,Voskresensky:1996ur}. 
However, a non-equilibrium single-particle distribution is not sufficient to describe correlations in the evolving system. As example, cluster formation in an expanding fireball requires the inclusion of higher order correlation functions to describe bound states like hadrons or nuclei.
Alternatively, the freeze-out concept assumes nuclear statistical equilibrium (NSE) during 
the expansion of the fireball, which is justified if the time $\tau_{\rm therm}$ for the relaxation to 
thermodynamic equilibrium is small compared to the variation time $\tau_{\rm exp}=q/\dot q$ 
of a parameter $q$ describing the thermodynamic state of the expanding system. 
The treatment of thermodynamic equilibrium is able to include all equilibrium correlations, in particular cluster formation. 
At freeze-out, $t = t_{\rm fo}$, collision processes which change the composition and the distribution die out. 
For $t>t_{\rm fo}$ baryon distributions evolve according to the mean-field description of the expansion. 
Note that at least under conditions of the LHC and highest RHIC energies for soft pions 
(with energies and momenta smaller than $m_\pi$)
 the time scales  characterizing elastic and inelastic (absorptive) processes are such that 
%$\tau_{\pi,{\rm el}}\ll \tau_{\pi,{\rm abs}}$ and  
$\tau_{\pi,{\rm el}}\ll t_{\rm fo}<\tau_{\pi,{\rm abs}}$. 
Thereby we may speak about the evolution of $\mu_\pi (t)$ till thermal freeze-out.
Although the expanding fireball approach with time dependent  temperature $T(t)$ 
and chemical potentials $\mu_c(t)$ for all hadrons including pions is rather reasonable it is just an approximation
to a more sophisticated many-body nonequilibrium approach given below. 

A systematic general approach to nonequilibrium is given by the NSO $\rho(t)$ which is the solution of the von Neumann equation with a given initial state \cite{Zubarev:1996}, 
\begin{equation}
\label{rho}
 \rho(t) = \lim_{\varepsilon \to 0} \varepsilon \int_{-\infty}^t dt' e^{\varepsilon (t - t')} U(t,t') \rho_{\rm rel}(t') U^\dagger(t,t')
\end{equation}
where for a hamiltonian $H$ which is not time-dependent holds 
$U(t,t')=\exp [(i/\hbar) H (t-t')]$, and the limit $\varepsilon \to 0$ has to be taken after the thermodynamic limit. 
Instead of a distribution $\rho_{\rm initial}(t_0)$ at an initial time $t_0$, according to Zubarev a relevant statistical operator 
$\rho_{\rm rel}(t')$
has been introduced that contains all relevant information of the system in the past $t'<t$. 
This relevant information characterizes the state of the system in non-equilibrium and will be discussed below.
The missing irrelevant correlations 
$\rho_{\rm irrel}(t)= \rho(t)-\rho_{\rm rel}(t)$ 
are assumed to be formed dynamically by the action of the time evolution operator $ U(t,t')$.

The nonequilibrium state of the system is characterized by observables $B_n$ in addition to the conserved observables such as total energy and particle numbers.
The relevant information is given by the averages $\langle B_n \rangle^t = {\rm Tr}\{ \rho(t) B_n\}$ of the relevant observables $B_n$.
The maximum of the information entropy ${\rm Tr}\{\rho_{\rm rel}(t) \ln \rho_{\rm rel}(t) \}$ at given averages is the generalized Gibbs distribution
\begin{equation}
\label{rhorel}
 \rho_{\rm rel}(t)= \frac{1}{Z_{\rm rel}(t)}e^{-\sum_n F_n(t) B_n};\qquad Z_{\rm rel}(t)={\rm Tr}e^{-\sum_n F_n(t) B_n}
\end{equation}
where the Lagrange multipliers $F_n(t)$ are determined by the self-consistency conditions
\begin{equation}
\label{sccond}
 \langle B_n \rangle^t_{\rm rel} = \langle B_n \rangle^t
\end{equation}
with $\langle B_n \rangle^t_{\rm rel} \equiv {\rm Tr}\{\rho_{\rm rel}(t) B_n \}$. 
With respect to the selection of the set of relevant observables  $B_n$ it should contain all conserved quantities which cannot be changed owing to the dynamics of the system. 
In an ergodic system, all other correlations are produced dynamically. 
We can include any set of observables to the relevant ones. 
In particular, if we include all slowly varying observables in the set of relevant observables $\{B_n\}$, we can expect that a shorter time is necessary to produce the missing non-equilibrium correlations.
This means that (\ref{rhorel}) gives already a good  approximation for $\rho(t)$ at finite $\varepsilon$ so that memory effects are less important. 

According to the NSO method, the equations of evolution (generalized kinetic equations) are obtained from
\begin{equation}
\label{kinB}
 \frac{d}{dt}\langle B_n \rangle^t= \lim_{\varepsilon \to 0} \frac{i \varepsilon}{\hbar} \int_{-\infty}^t dt'e^{\epsilon (t'-t)} {\rm Tr}
\left\{\rho_{\rm rel}(t')e^{iH(t'-t)/\hbar}[H,B_n]e^{iH(t-t')/\hbar}\right\}~,
\end{equation}
where we inserted the time derivative of the NSO (\ref{rho}) and used the self-consistency conditions (\ref{sccond}). 
%In the case that $B_n$ do not belong to the set of relevant observables a more general expression results because the self-consistency relation (\ref{sccond}) cannot be used.

The correct reproduction of the relevant information in the past gives the possibility to form the irrelevant correlations very fast so that a perturbation expansion is possible. 
Although the expression (\ref{rho}) is correct for any choice of relevant observables after performing the limit 
$\varepsilon \to 0$, an appropriate choice of the set of relevant observables $\{ B_n \}$ allows us to make expansions 
quickly convergent so that, for instance, the Markov approximation can be performed\footnote{We speak here about memory effects associated with dying initial correlations. There are memory effects related to processes  described by diagrams with more than two vertices in the non-equilibrium Greens function technique. These effects can be neglected  only for dilute gases but in general they are  important. They result in the famous  $T^2\ln T$ correction to the specific heat of $^3$He, cf. 
\cite{Carneiro:1975,Ivanov:1999tj}. Since this correction is quite comparable (numerically) to the leading term in the specific heat $ (\propto T )$, one may claim that liquid  $^3$He is a liquid with quite strong memory effects from the point of view of kinetics.}.
In Sec.~\ref{sec:NSO} we discuss this issue in detail. It is our main goal to show that the optimal path for the nonequilibrium evolution (i.e. a sufficient broad choice of relevant observables $B_n$) must be found to provide us with a precise description already in low orders of perturbation theory.

A correct description of the evolution is also necessary for the expanding fireball if the freeze-out approximation is used. 
We cannot assume that the system at freeze-out is strictly in thermodynamic equilibrium.
A simple relaxation-time ansatz is not sufficient but a more detailed description of the time evolution is necessary. 
An important feature of the evolution is that a perturbation expansion in powers of the interaction which is often used cannot give the formation of bound states or quantum condensates in any finite order of perturbation theory.
The introduction of the relevant NSO provides us with the description of those phenomena. 
In particular we show that the formation of an intermediate pion Bose-Einstein condensate can be described in our approach.   
It is an advantage of the Zubarev approach that initial correlations are included via the relevant statistical operator 
$\rho_{\rm rel}(t)$, in contrast to the kinetic theory which is based on the single-particle distribution function.

\section{Model for pions in heavy-ion collisions  at ultra high energies}

To explain our approach denoted as the NSO method, we discuss the pion production 
in heavy-ion collisions. After an initial stage where hadrons are formed, we consider a fireball consisting of pions, nucleons $N$, 
$\Delta$ resonances, and further hadronic states such as $N^*$ and $K$ mesons containing strangeness degrees of freedom.
In occupation number representation, $a^+_{{\bf p}, c}, a_{{\bf p}, c}$ are the creation/annihilation operators of a particle 
in the quantum state ${\bf p}, c$ given by the species $c$ (including spin) and momentum $\bf p$. 
The Hamiltonian $H=H^0+H'$ contains the  kinetic part 
\begin{equation}
\label{H0}
H^0 = \sum_{{\bf p},c} \sqrt{p^2 + m_c^2} a^+_{{\bf p}, c} a_{{\bf p}, c}= \sum_{{\bf p},c} E_{{\bf p}, c} \,a^+_{{\bf p}, c} a_{{\bf p}, c}
\end{equation}
and the interaction part $H'$. 
At highest RHIC and LHC energies the fireball is dominated by pions. Approximately one may speak of a pure pion gas.
So further within  our model, we consider only special processes, which concern the pion distribution, $c$ runs over pion species. 
We split the interaction into an elastic part describing collisions which conserve
the pion number, $H_{\rm col}$, and an inelastic part describing reactions, $H_{\rm reac}$, where the pion number is 
changed, $H'=H_{\rm col}+H_{\rm reac}$.

Expressions for the meson - meson 
%and meson - nucleon 
interaction are found in the literature 
\cite{Welke,Gerber:1990yb}.
$\pi - \pi$ interactions at 
%low 
high energies 
%(below 1 GeV) 
are predominantly elastic 
%\cite{Gavin:1991ki}, 
implying that at low density the number of pions is effectively conserved.
We assume elastic $\pi - \pi$ 
%and $\pi - N$ 
scattering of the form 
\begin{equation}
\label{Hcol}
H_{\rm col} = \frac{1}{2} \sum_{{\bf p_1, p_2, p'_1, p'_2}; c,d}  \lambda_{c,d}({\bf p_1,p_2;p'_{1},p'_{2}})
 a_{{\bf p_1},c}^\dagger a_{{\bf p_2},d}^\dagger a_{{\bf p'_2},d}a_{{\bf p'_1},c}.
\end{equation}
We assume that at $ t<t_{\pi,{\rm cfo}}$ a state overpopulated by soft pions is formed, for $ t>t_{\pi,{\rm cfo}}$
the collisions conserve the particle number, but evolve the distribution function to a thermal equilibrium distribution with a corresponding short relaxation time $\tau_{\rm col}$.
These collision processes may be happen via a virtual states such as the $\rho$ and $\sigma$ mesons or other resonances.
We note that the matrix element $ \lambda_{\pi,\pi}({\bf p_1,p_2;p'_{1},p'_{2}})$ of the $\pi-\pi$ interaction can be taken in a separable form \cite{Aouissat:1991tb,Barz:1992ra}
so that the Bethe-Goldstone equation for the T-matrix of the $\pi-\pi$ scattering in the pion gas can be 
solved straightforwardly, resulting in in-medium scattering phase shifts and cross sections with resonances 
\cite{Barz:1992ra,Barz:1992ic,Expansion}
as well as the corresponding equation of state \cite{Rapp:1995ir,Rapp:1995py}.
%in the Beth-Uhlenbeck approach. 

Interactions of pions with baryons can also be described as particle number conserving $2\to2$ processes with a 
Hamiltonian of the form (\ref{Hcol}). 
See for example the recent work \cite{Kamano:2019gtm} on the ANL-Osaka model which provides an excellent description 
of existing $\pi - N$ scattering data.
These processes contain, in particular, the $\Delta$ and $N$ resonances that become very important for heavy-ion collision experiments with lower c.m. energies at SPS, SIS-18 and the future FAIR and NICA experiments. 
For our discussion of results from the LHC experiment in the present work the processes involving baryons are not important and will not be treated explicitly here.

If the particle number is fixed, neglecting processes described by $H_{\rm reac}$, a pion gas in thermodynamic equilibrium may form a Bose-Einstein condensate at high phase space occupation densities and sufficiently low temperatures.
It is possible that the expanding fireball will meet such parameter values of the pion phase space during the nonequilibrium evolution.

There are also processes which change the particle numbers of the different species which contribute to the interaction part 
of the Hamiltonian $H_{\rm reac}$. 
As example we have $\pi+\pi \rightleftharpoons 4 \pi$ \cite{Voskresensky:1995tx} or the formation of other mesons such as 
$\pi+\pi \to \bar K +K$ which decay in other channels, see Lin and Ko \cite{Welke}. 
As shown there, because of the threshold for these reactions and the small cross sections compared to the elastic collisions the corresponding relaxation time $\tau_{\rm reac}$ to establish chemical equilibrium is large. 
This is a slow process not of relevance for the time scales considered here.
Other reactions involving resonant correlations such as $\Delta \rightleftharpoons N +\pi$ contribute to collision processes via virtual states and conserve the particle number, but have also a small branching ratio for number non-conserving processes.
These reactive collisions which change the pion number are assumed to be weak in comparison to the quasielastic collisions and can be discarded for the short-time evolution, but are relevant for the evolution on long time scales to produce chemical equilibrium.

The $\pi - \pi$ collision term was treated in Boltzmann equation calculations by Welke and Bertsch 
\cite{Welke}. 
Strong pion interaction processes which change the pion number have been considered, e.g., in 
Ref.~\cite{Voskresensky:1995tx}.

Also other work has been performed using transport codes to describe the time evolution of the pion distribution function and to solve the low-$p_T$ enhancement puzzle. We discuss here the more general 
NSO approach to give an approach which goes beyond the single-particle distribution function considered 
in the transport codes which are based on kinetic equations.

\section{The relevant statistical operator}
\label{sec:NSO}

Our main point is the selection of the set of relevant observables $B_n$ which determines the convergence and the accuracy of the non-equilibrium description.  
We discuss three examples, the Kubo case considering only conserved quantities, the kinetic theory considering the single-particle occupation numbers, and the formation of a condensate where amplitudes are added. 
In principle, all three choices for the set of relevant observables should give the same results if the limit $\varepsilon \to 0$ is correctly performed. 
However, because we use perturbation expansions and Fermis Golden rule, these approximations lead to different results.

\subsection{Kubo case}

Within the NSO method, there is no prescription for the choice of the set $\{ B_n\}$ of relevant observables. 
Only conserved observables have to be included because their averages cannot be changed dynamically.

A minimum set of relevant observables of the pion-nucleon system (Kubo case) 
is the energy $H$ which is conserved.
The number of pions is not strictly conserved. 
Because the pion number is not prescribed, in equilibrium no corresponding chemical potential $\mu_\pi$ can be introduced. 
Formally, one takes $\mu_\pi=0$ similar to the photon system. 
Thus in Kubo case one supposes that $\tau_{\pi,{\rm abs}}\ll t_{\rm fo}$ (contrary to the case studied by us in this paper).
The number of baryons $N_b$ is conserved. 
Also the charge has to be considered as conserved quantity.
With this selection of relevant observables, we obtain from the maximum principle for the relevant entropy 
the grand canonical distribution
\begin{equation}
\label{relrho}
 \rho^{(0)}_{\rm rel}(t)= \frac{1}{Z^{(0)}_{\rm rel}(t)}e^{- \beta(t) (H-\sum_\alpha \mu_\alpha(t) N_\alpha)}~.
\end{equation}

Because the system is expanding, the thermodynamic averages are also changing with time and also 
the corresponding Lagrange parameters. We can adopt the blast wave model 
\cite{Siemens:1978pb,blast,blast2,blast3,Florkowski:2010zz} to describe 
the expansion of the fireball. If we assume that the velocity is proportional to the distance from the center, 
the density is decreasing with time. Assuming adiabatic expansion, the entropy is constant, 
but the temperature is also decreasing. This hydrodynamical model may serve as an approximation 
to describe the time dependence of the average density and energy and, according to the self-consistency conditions (\ref{sccond}), $\mu_\alpha(t)$ and $\beta(t)$. 

Starting from a nonequilibrium state, we consider the relaxation to the local thermodynamic equilibrium.
The time behavior of the observables $N_c$ and $H^{\rm MF}_c$ of the particle number of the species $c$ 
and its mean-field energy (see below) is given by 
\begin{equation}
\label{dNc}
 \frac{d}{dt}\langle N_c \rangle^t= \frac{i}{\hbar}\langle [H,N_c] \rangle^t_{\rm rel} -\frac{1}{\hbar^2}\int_{-\infty}^0 dt'
 e^{\epsilon t'} 
 {\rm Tr} \{ \rho_{\rm rel}(t)[H(t'),[H, N_c ]] \}~,
\end{equation}

\begin{equation}
\label{dkin}
 \frac{d}{dt}\langle H^{\rm MF}_c \rangle^t=\frac{i}{\hbar}\langle [H,H^{\rm MF}_c] \rangle^t_{\rm rel} -\frac{1}{\hbar^2}
 \int_{-\infty}^0 dt' e^{\epsilon t'} 
 {\rm Tr} \{ \rho_{\rm rel}(t)[H(t'),[H, H^{\rm MF}_c ]] \}~.
\end{equation}

Evaluating the correlation functions in Born approximation for the pion system, we observe a behavior
different from the one that would be consistent with the assumption of $\mu_\pi=0$. 
Because the particle numbers $N_\pi$ are conserved with respect to the elastic collisions $H_{\rm col}$,
only the inelastic collisions $H_{\rm reac}$ contribute. 
This makes the time derivative (\ref{dNc}) small.
In contrast, the thermalization process (\ref{dkin}) is dominated by $H_{\rm col}$ so that the exchange of 
energy between the different components $c$ and the momentum states is a fast process.
 
We can calculate the corresponding relaxation times 
$\tau^{-1}_i = -[d\langle B_i \rangle^t/dt]/\langle B_i \rangle^t$ for the observables 
of local thermodynamic equilibrium and compare it with the expansion time scale
$\tau^{-1}_{\rm exp} = \partial_\mu u^\mu \approx -n_b^{-1} [d n_b/dt]$. 
The freeze-out time $t_{\rm fo}$ is given by the condition that the increasing $\tau_i(t)$ 
becomes equal to $\tau_{\rm exp}$.

It is evident that this is a very global approach. We cannot assume that at any freeze-out time $t_{\rm fo}$
the system is well approximated by the equilibrium distribution (\ref{relrho}). 
A more detailed description of the nonequilibrium state is necessary, in particular if there exist 
long-living correlations. Indeed, the relaxation to thermodynamic equilibrium (\ref{relrho}) implies also the 
achievement of the total pion number in equilibrium which is determined only by $T$ and $\mu_\pi=0$ in thermodynamic equilibrium. 
Because the processes which change the pion number are weak under conditions at RHIC and LHC which we focus on, 
the corresponding relaxation times are long and the Kubo case is not valid for our considerations.
To have an appropriate description of the nonequilibrium process, these slow modes should be included in $\rho_{\rm rel}(t)$.
At freeze-out we therefore expect that not the thermodynamic equilibrium but a more general nonequilibrium distribution is seen.

\subsection{Pion number as relevant observable}

Long-living correlations have to be implemented in the set of relevant observables to 
improve the convergence of the perturbation expansion, and to apply the Markov approximation.
For the pion
%pion-nucleon 
system considered here, we have elastic collisions which conserve the particle numbers
and in general inelastic reactions where the particle numbers of the constituents $c$ are changed.
Because the conserving interaction $H_{\rm col}$ leads to cross sections which are large
compared with cross sections for the non-conserving  interaction $H_{\rm reac}$, the pion number
$N_\pi$ is an observable which changes slowly with time and should be included to the set of
relevant observables so that the index $\alpha$ in Eq. (\ref{relrho}) goes over all pion species $c$. 
The new condition 
\begin{equation}
\label{scpi}
 \langle N_\pi \rangle^t_{\rm rel} = \langle N_\pi \rangle^t
\end{equation}
is not given by the external condition of the expanding fireball but must be calculated self-consistently
solving the corresponding equation of evolution (\ref{dNc}).

A new feature of the relevant distribution including the pion number is the possibility of a singularity
when the self-consistency conditions (\ref{sccond}) are solved. As well known from the ideal Bose gas,
we have to treat the occupation of the ground state separately so that below a critical temperature we have
$\langle N_\pi \rangle=\langle N^{\rm norm}_\pi \rangle+\langle N^{\rm cond}_\pi \rangle$ with 
the normal component $N^{\rm norm}_\pi=\sum_{p>0} a^+_{{\bf p}, \pi} a_{{\bf p}, \pi}$ and the condensate
$N^{\rm cond}_\pi= a^+_{0, \pi} a_{0, \pi}$. The corresponding relevant operator reads
\begin{equation}
\label{relBose}
 \rho^{\rm Bose}_{\rm rel}(t)= \frac{1}{Z^{\rm Bose}_{\rm rel}(t)}
e^{- \beta(t) [H-\sum_c \mu_c(t) N^{\rm norm}_c]-F_{0, \pi}(t)  a^+_{0, \pi} a_{0, \pi} }~.
\end{equation}
The new Lagrange parameter $F_{0, \pi}(t)$ follows from the self-consistency relation (\ref{scpi}) in perturbation expansion with respect to $H'$ as 
\begin{equation}
\label{F0}
 \langle N^{\rm cond}_\pi \rangle^t_{\rm rel} = \frac{1}{e^{\beta(t)[E_{0, \pi}-\mu_\pi(t)]+F_{0, \pi}(t)}-1}~,
\end{equation}
which is a macroscopic number if the temperature is below the critical one.
The normal component $\langle N^{\rm norm}_\pi \rangle^t_{\rm rel}$ is only a function of $\beta(t)$ as well known; 
we have $E_{0, \pi}-\mu_\pi(t)=0$ in the condensate state. 
Then, the Bose condensate $\langle N^{\rm cond}_\pi \rangle^t_{\rm rel}$ is a macroscopic number so that $F_{0, \pi}(t)$ is infinitesimal small. 
Below we improve this shortcoming introducing coherent states. 

We can also solve the evolution equations (\ref{dNc}) with the relevant statistical operator (\ref{relBose}). The corresponding relaxation times are given by the interaction part $H'$. However,
they are very different. As before, the elastic collisions thermalize the kinetic energies of the 
components (\ref{dkin}) leading to the relaxation time $\tau_{\rm therm}$.

To describe chemical relaxation (\ref{dNc}) we can simplify the von Neumann equation as 
\begin{equation}
 \frac{\partial}{\partial t}\rho(t)=\frac{1}{i \hbar}\left[(H^0+H_{\rm reac}),\rho(t)\right]
-\frac{1}{\tau_{\rm therm}} \left(\rho(t)-\rho^{\rm Bose}_{\rm rel}(t) \right).
\end{equation}
This is possible if the thermalization is very fast compared to the formation of the chemical equilibrium.
The solution is given by Eq. (\ref{rho}) after replacing $\epsilon$ by $1/\tau_{\rm therm}$ and $H'$ by 
$H_{\rm reac}$.

With respect to the evolution of the fireball, we have the result that full thermodynamic equilibrium must not 
necessarily occur at the freeze-out time. The reaction rates become small during the expansion so that
the relevant distribution (\ref{relBose}) at $t_{\rm fo}$ is seen. 
In contrast to the thermodynamic equilibrium (\ref{relrho}), a Bose-Einstein condensate of pions is possible. 

\subsection{Kinetic equations}

We can further improve the relevant statistical operator considering the occupation numbers $n_{{\bf p},c}= a^+_{{\bf p},c} a_{{\bf p},c}$ of 
the single-particle states as relevant observables. Formally, instead of $N_c$ each 
single-particle state is taken similar to a new species.
The time dependence of the mean occupation of this state leads to the kinetic equations.
For details of the derivation see, e.g., Ref.~\cite{Roepke2013}, Eq.~(4.100).

We consider the time evolution of the pion distribution function $\langle n_{{\bf p},c} \rangle^t$ as the diagonal part of the Wigner distribution function \cite{Roepke2013}.
The relevant statistical operator has the form
\begin{equation}
\label{relrhokin}
 \rho^{\rm kin}_{\rm rel}(t)= \frac{1}{Z^{\rm kin}_{\rm rel}(t)}e^{-\sum_{{\bf p},c} s_{{\bf p},c}(t) a^+_{{\bf p},c} a_{{\bf p},c}}
\end{equation}
with the corresponding self-consistency conditions to eliminate the Lagrange parameters $s_{{\bf p},c}(t)$,
\begin{equation}
 \langle n_{{\bf p},c} \rangle^t=\frac{1}{e^{s_{{\bf p},c}(t)}-1}.
\end{equation}
For the time evolution the following kinetic equation is obtained from (\ref{kinB}) after integration by parts and using (\ref{sccond}), see also \cite{Roepke2013}
\begin{equation}
\frac{d}{dt}\langle n_{{\bf p},c} \rangle^t = \frac{1}{\hbar^2}\int_{-\infty}^0 dt'e^{\epsilon t'}{\rm Tr}\left\{[H,n_{{\bf p},c}]
e^{(i/\hbar)Ht'}[H,\rho^{\rm kin}_{\rm rel}(t)]e^{-(i/\hbar)Ht'}\right\}
\end{equation}
if we neglect the explicit time dependence of $ \rho^{\rm kin}_{\rm rel}(t)$.
In the approximation of binary collisions we get the quantum statistical Boltzmann equation
\begin{eqnarray}
\label{BEq}
 \frac{d}{dt}\langle n_{{\bf p}_1} \rangle^t_{\rm coll}&=&\frac{2\pi}{\hbar}\sum_{{\bf p}_2,{\bf p}'_1,{\bf p}'_2}
\delta(E_{{\bf p}_1}+E_{{\bf p}_2}-E_{{\bf p}_1'}-E_{{\bf p}_2'})\delta_{{\bf p}_1+{\bf p}_2-{\bf p}_1'-{\bf p}_2'}
|t({\bf p}_1{\bf p}_2,{\bf p}_1'{\bf p}_2')+t({\bf p}_1{\bf p}_2,{\bf p}_2'{\bf p}_1')|^2 \nonumber \\
&&\times \left\{\langle n_{{\bf p}_1'} \rangle^t \langle n_{{\bf p}_2'} \rangle^t[1+\langle n_{{\bf p}_1} \rangle^t][1+\langle n_{{\bf p}_2} \rangle^t]
-\langle n_{{\bf p}_1} \rangle^t \langle n_{{\bf p}_2} \rangle^t[1+\langle n_{{\bf p}_1'} \rangle^t][1+\langle n_{{\bf p}'} \rangle^t] \right\}
\end{eqnarray}
where the two-particle T-matrix is given by the interaction potential in Born approximation \cite{Roepke2013}.
Near thermodynamic equilibrium,
\begin{equation}
\langle n_{{\bf p},c} \rangle_{\rm eq}=\frac{1}{e^{E_{{\bf p},c}/T-\mu_c/T}-1}~,
\end{equation}
we can approximate the Boltzmann equation in relaxation time approximation as:
\begin{equation}
\frac{d}{dt}\langle n_{{\bf p},c} \rangle^t = -\frac{1}{\tau_{{\bf p},c}} (\langle n_{{\bf p},c} \rangle^t -\langle n_{{\bf p},c} \rangle_{\rm eq})~,
\end{equation}
where the relaxation time $\tau_{{\bf p},c}$ is calculated from a microscopic collision process. 
This approach of a relaxation time $\tau_{\rm col}$ of collisions as the thermal average over $\tau_{{\bf p},c}$ 
\cite{Roepke2013} or the corresponding collision frequency is used in the relevant literature (see, e.g., \cite{Welke}).
Note that the relaxation time ansatz (19) with $\bf p$-dependent relaxation time does not obey, in general, the conservation of particle number. 
According to Mermin \cite{Mermin:1970zz}, this defect is removed if the relaxation occurs not to the equilibrium state but to a relevant operator which accounts for the conservation of particle number, see also \cite{Selchow:2002}.

Thermal freeze-out is obtained at the time $t_{\rm fo}$ when $\tau_{\rm col}(t_{\rm fo}) = \tau_{\rm exp}(t_{\rm fo})$, where the scattering time scale for a given particle species $c$, working in favor of equilibration, can be computed locally from the local 
densities $n_d$, thermal (relative) velocities $v_{cd}$, and total scattering cross sections $\sigma_{cd}$ between the particles $c$ and $d$ \cite{blast,blast2,blast3} after momentum average
\begin{equation}
 \frac{1}{\tau_{\rm col}}=\sum_d \langle v_{cd} \sigma_{cd} \rangle n_d ~.
\end{equation}

\subsection{Nonequilibrium state with condensate formation}

Alternatively, we can construct another  relevant operator with the single-particle occupation numbers $n_{{\bf p},c}$, 
but containing also non-diagonal parts and, in addition, also single construction operators $a_{{\bf p},c}$ and $a^+_{{\bf p},c}$. Corresponding expressions are known from the theory of coherent states which are of interest to describe Bose-Einstein condensates. We can construct the relevant entropy with arbitrary powers of the creation and annihilation operators, corresponding to a very general expansion of the entropy operator in occupation number representation. 

As a simple case, we construct the relevant statistical operator 
\begin{equation}
\label{relrhocon}
 \rho^{\rm coh}_{\rm rel}(t)= \frac{1}{Z^{\rm coh}_{\rm rel}(t)}e^{\sum_{{\bf p},c} [F^*_{{\bf p},c}(t) a_{{\bf p},c} +F_{{\bf p},c}(t) a^+_{{\bf p},c} 
 -s_{{\bf p},c}(t) a^+_{{\bf p},c} a_{{\bf p},c}]} 
%-\sum_c \beta_c(t) (H_c-\mu_c(t) N_c)}
=e^{-S^{(2)}(t)}
\end{equation}
with the corresponding expression for the partition function $Z^{\rm coh}_{\rm rel}(t)$. Higher order contributions 
such as $ a^+_{{\bf p},c} a^+_{{\bf p},c}$ are also possible, as well as non-diagonal terms (describing systems which are not homogeneous in space) but will not be considered here. A similar approach has been used 
for superfluidity in strongly coupled fermion systems \cite{rsuperfl,rsuperfl2}. Note that this bilinear form of the entropy $S^{(2)}(t)$ may be extended including higher than second order terms in $a^+_{{\bf p},c}$ and $a_{{\bf p},c}$. This is necessary to describe, e.g., total energy conservation or
the formation of bound states.

We have to eliminate the Lagrange multipliers $F^*_{{\bf p},c}(t), F_{{\bf p},c}(t), s_{{\bf p},c}(t)$ using the self-consistency conditions (\ref{sccond}). 
The evaluation of correlation functions becomes quite simple if the relevant statistical operator 
(\ref{relrho}) is diagonal in the occupation number representation. 
We transform $a_{{\bf p},c}=b_{{\bf p},c}+B_{{\bf p},c}(t)$ where $b_{{\bf p},c}$ obey the usual commutation relations  for bosons, 
and $B_{{\bf p},c}(t)=F_{{\bf p},c}(t)/s_{{\bf p},c}(t)$ is a c-number. We obtain the diagonal form 
\begin{equation}
\label{Sbil}
S^{(2)}(t)= \sum_{{\bf p},c}[s_{{\bf p},c}(t) b^+_{{\bf p},c} b_{{\bf p},c}-|F_{{\bf p},c}(t)|^2/s_{{\bf p},c}(t)]
\end{equation}
for the bilinear relevant entropy $S^{(2)}(t)$, the c-number term can be canceled with $ Z^{\rm coh}_{\rm rel}(t)$.
Then, the evaluation of $\langle n_{{\bf p},c} \rangle_{\rm rel}$ is quite simple and yields the well-known result
\begin{equation}
\label{FBose}
\langle b^+_{{\bf p},c} b_{{\bf p},c} \rangle_{\rm rel}^t= \frac{1}{e^{s_{{\bf p},c}(t) } - 1}=f_{{\bf p},c}(t),\qquad 
\langle b^+_{{\bf p},c} \rangle_{\rm rel}^t=\langle b_{{\bf p},c} \rangle_{\rm rel}^t=0
\end{equation}
what is the Bose distribution for the ideal quantum gas, but with nonequilibrium parameter $s_{{\bf p},c}(t)\ge 0$ which are 
determined by the given averages.
The mean occupation numbers follow as $\langle n_{{\bf p},c} \rangle_{\rm rel}^t=[e^{s_{{\bf p},c}(t) } - 1]^{-1}
+|B_{{\bf p},c}(t)|^2$. In addition, we find $\langle a_{{\bf p},c} \rangle_{\rm rel}^t= B_{{\bf p},c}(t)$. With these relations, the Lagrange parameters in Eq. (\ref{relrhocon}) can be eliminated.

As before, the dynamics of the many-particle system is described by the Hamiltonian $H=H^{0}+H_{\rm col}+H_{\rm reac}$,
defined in Eqs. (\ref{H0}), (\ref{Hcol}).
We extract the mean-field terms (MF) from the interaction ($1 = \{{\bf p},c\}$)
\begin{equation}
 H=\sum_1 E^{\rm MF}(1,t) a_1^+a_1+ \frac{1}{2} \sum_{12} \Delta_{\rm pair}^{\rm MF}(12,t) a_1^+a^+_2+{\rm c.c.}
+ \frac{1}{2} \sum_{121'2'}V(12,1'2') a_1^+a^+_2a_{2'}a_{1'} -{\rm (MF)}
\end{equation}
with  $E^{\rm MF}(1,t)=E(1)+\sum_2V(12,12)_{\rm ex}\langle n_2 \rangle^t$ and 
$\Delta_{\rm pair}^{\rm MF}(12,t)=\sum_{1'2'} V(12,1'2') \langle a_{2'}a_{1'} \rangle^t$.
We will not consider pairing so that $\Delta_{\rm pair}^{\rm MF}(12,t)=0$. 
In the case of fermions pairing was considered in Ref. \cite{rsuperfl,rsuperfl2}, 
which can also transformed to the bilinear form (\ref{Sbil}) applying the Bogoliubov transformation. 
In the case of a Bose gas considered here, the mean-field terms contain also averages 
$\langle a^+_2 a_{2'} a_{1'}\rangle^t$ of the condensate mode so that
\begin{equation}
H^{\rm MF}(t)=\sum_1 E^{\rm MF}(1,t) a^+_1a_1+\frac{1}{2} \Delta_{\rm cond}^{\rm MF}(1,t) a_1^++{\rm c.c.}
\end{equation}
with $\Delta_{\rm cond}^{\rm MF}(1,t)=\sum_{21'2'}V(12,1'2') \langle a^+_2 a_{2'} a_{1'}\rangle^t$.

According to the NSO method, the kinetic equations are obtained from the equations of evolution (\ref{kinB}) for the relevant observables
\begin{equation}
 \frac{d}{dt}\langle B_n \rangle^t= \lim_{\epsilon \to 0} \frac{i \epsilon}{\hbar} \int_{-\infty}^t dt'e^{\epsilon (t'-t)} {\rm Tr}
\left\{e^{-S^{(2)}(t')}e^{iH(t'-t)/\hbar}[H,B_n]e^{-iH(t-t')/\hbar}\right\}.
\end{equation}
We apply perturbation theory with respect to the deviation $\Delta H$ from the mean-field expression which can be incorporated into 
$s_{{\bf p},c}(t) =\beta_c(t)[E^{\rm MF}_{{\bf p},c}(t)-\mu_c(t)]+\delta f_{{\bf p},c}(t) =f^{\rm MF}_{{\bf p},c}(t)+\delta f_{{\bf p},c}(t)$. 
The new Lagrange parameter $\beta_c(t),\mu_c(t)$
are introduced to describe the total particle number $N_c$ and mean-field energy $H_c^{\rm MF}$ of the species $c$. 
The perturbation expansion is performed with respect to $\Delta S (t)=S^{(2)}(t)-S^0(t)$ with 
$S^0(t)=\beta(t)[H^{\rm MF}(t)-\sum_c \mu_c(t) N_c]$ 
where $\beta(t)$ is determined by the average of the total energy $H$.
We have ($S^{(2)}(t)$ commutes with $S^0(t)$ in lowest order of perturbation theory)
\begin{equation}
 \frac{d}{dt}\langle a^+_1 \rangle^t= \lim_{\epsilon \to 0} \frac{i \epsilon}{\hbar} \int_{-\infty}^0 dt'
e^{\epsilon t'} e^{-i\mu_1t'/\hbar}  {\rm Tr}
\left\{e^{-S^{(2)}(t'+t)}\left[E(1)a^+_1+\frac{1}{2}\sum_{1'22'} V(1'2',12)a^+_{1'}a^+_{2'}a_2\right]\right\}
\end{equation}
and the corresponding equations of evolution for the other relevant observables $N_c, H_c^{\rm MF},a_{{\bf p},c}$.
To evaluate the trace we perform the transformation of the relevant statistical operator 
(\ref{relrho}) to the diagonal form (\ref{Sbil}) and have
\begin{equation}
 \frac{d}{dt}\langle a^+_1 \rangle^t= \frac{i}{\hbar} E^{\rm MF}(1,t)\lim_{\epsilon \to 0}  \epsilon \int_{-\infty}^0 dt'
e^{\epsilon t'} e^{-i\mu_1t'/\hbar} F^*_1(t+t'),
\end{equation}
where it was assumed that the mean-field energy $E^{\rm MF}(1,t)=E(1)+\sum_2V(12,12)_{\rm ex}  f_{2}(t) $ depends only weakly on time so that it can be extracted from the integral which is determined by the collisions.
In addition, we suppose that a condensate mode is only in the state ${\bf p}_1,c_1$ and $V(11,11)=0$ 
with $1$ denoting the state of lowest mean field energy.
In the stationary state we assume a periodic dependence on time, $F_1(t)=F_1^0 e^{i \omega t}$. 
Then, the integral can be performed with the result $\omega =\mu_1/\hbar$. 
The amplitude $\langle a_1 \rangle^t=F(t)$ depends periodically on time. 
We obtain the condition $\hbar \omega =\mu_1=E^{\rm MF}(1)$ for a stationary solution, considering the lowest order 
(mean-field approximation). 
Similar results hold for $ \langle a^+_1 \rangle^t $.
For the other relevant observables, the time derivative vanishes in lowest order of perturbation expansion with respect of the interaction. 

To obtain the evolution of the averages one has to consider higher orders of the interaction.
It is convenient to use the following expression for the NSO obtained from (\ref{rho}) 
after integration by parts
\begin{equation}
 \rho(t)=\rho_{\rm rel}(t)-\lim_{\epsilon\to 0}\int_{-\infty}^t e^{\epsilon (t'-t)} U(t,t')\left\{\frac{i}{\hbar}[H,\rho_{\rm rel}(t')]+\frac{\partial}{\partial t'}\rho_{\rm rel}(t')
\right\} U(t',t) dt'.
\end{equation}
so that for the averages
\begin{eqnarray}
&&\label{kinB1}
 \frac{d}{dt}\langle B_n \rangle^t= \frac{i}{\hbar} {\rm Tr}\{\rho_{\rm rel}(t) [H,B_n]\}\nonumber\\&&
+\frac{1}{\hbar^2} \int_{-\infty}^0 dt'e^{\epsilon t'} {\rm Tr}
\left\{[H,B_n]e^{iHt'/\hbar}\left([H,\rho_{\rm rel}(t')]+\frac{\hbar \partial}{i\partial t'}\rho_{\rm rel}(t')\right)e^{-iHt'/\hbar}\right\}~.
\end{eqnarray}
In Markov approximation, the time dependence of $\rho_{\rm rel}(t')$ is neglected, we have
the Boltzmann-like form of the equations of evolution,
see also (\ref{dNc}), (\ref{dkin}) which is obtained after cyclic permutation within the trace. 
In our case, we cannot 
assume that the time dependence of $F_1(t)\propto \exp(i\omega t)$ is slow, only after the transformation to the diagonal 
form the remaining $s_1(t)$ may be slow.

Using Wick's theorem, the evaluation of the first term of the r.h.s. of (\ref{kinB1}) is immediately 
done if we transform to the diagonal form of the relevant statistical operator. 
The second term describes the collision between the pions and gives the relaxation to the 
intermediate relevant state showing the condensate distribution.
The evaluation of the time dependence of occupation numbers $\langle n_{{\bf p},c}\rangle^t$ coincides with the expression (\ref{BEq}) but replacing $\langle n_{{\bf p},c}\rangle^t$ by
$ f_{{\bf p},c}(t)+|B_{{\bf p},c}(t)|^2$.

The time evolution of the amplitude follows in Born approximation as 
\begin{eqnarray}
\label{BEqcond}
\frac{d}{dt}\langle a^+_1 \rangle^t&=&\frac{i}{\hbar}E_1^{\rm MF} B^*_1(t)+\frac{\pi}{2\hbar}B^*_1(t)
 \sum_{1'22'}|V_{\rm ex}(12,1'2')|^2
\delta(E_{{\bf p}_1}+E_{{\bf p}_2}-E_{{\bf p}_1'}-E_{{\bf p}_2'})\delta_{{\bf p}_1+{\bf p}_2-{\bf p}_1'-{\bf p}_2'}\nonumber \\
&&\times \left\{f_{1'} f_{2'}(1+f_2)-f_2(1+f_{2'})(1+f_{1'}) \right\}
\end{eqnarray}

Stationary solution is the grand canonical distribution with the chemical potential given by the pion number density.
If the pion chemical potential approaches the lowest pion energy state, a quantum condensates will be formed which
is described by a coherent state. The time evolution of the condensate is characterized by the collision time 
$\tau_{\rm cond}$.

\subsection{Quantum master equation}

The nonequilibrium evolution of the pion system can also be treated considering it as an open system 
coupled to a bath. 
We can consider the gluon system as the bath in the stage of evolution where the pion are formed from the hot quark-gluon plasma,
or we can consider the interaction between pions (e.g. via $\rho$ mesons) as a bath.
We also can consider the pion Bose-Einstein condensate as the relevant subsystem interacting with the normal pion gas.
A quantum master equation is derived which contains a Lindblad term \cite{Zubarev:1996},
and a solution can be performed using coherent states. We will not discuss this interesting approach here
but mention that the treatment of open many-particle systems is also possible within the Zubarev NSO method,
leading to quantum master equations \cite{Zubarev:1996}. 
See for instance Ref.~\cite{Akamatsu:2014qsa} for the treatment of heavy quarkonia kinetics in a quark-gluon plasma.

%%%%%%%%%%%%%%%%%%%%%%%%%%%%%%%%%%%%%%%%%%%%%%%%%%%%
\section{Discussion}
\label{sec:results}	

We refer to the central 200 AGeV $^{16}$O+Au data of the NA35 Collaboration, see  \cite{Welke}. 
For more recent data see also the discussion of Ref.~\cite{Begun:2015ifa}.
Assuming hadronization (formation time) at $t_0\sim 1$ fm/c and temperature of about 160 MeV, the equilibrium pion density at $\mu_\pi=0$ is $n_{\pi, {\rm normal}}\approx 0.15$ fm$^{-3}$, in contrast to the observed density $n_\pi (t_0) \sim 1$  fm$^{-3}$.
This motivates to consider a strong macroscopic occupation of the lowest momentum state, forming a pion Bose-Einstein condensate.

We use the hydrodynamical expansion under conservation of the initial entropy $S_0$ which determines 
the temperature evolution according to $s(T) V(\tau) = S_0 = {\rm const.}$, where for the entropy density we use a fit to lattice QCD data from Ref.~\cite{Pena:2013hd}.

For the expansion of the fireball we can adopt a Bjorken like picture where in the first stage ($t<10$ fm/c) we have one-dimensional expansion with $n_\pi(t)\sim n_\pi(t_0) t_0/t$, and afterwards three-dimensional expansion where 
$n_\pi(t)\propto t^{-3}$. 
The expansion rate $\tau_{\rm exp}^{-1} = |\dot{n}|/n \sim 1/t$ drops down as $1/t$.

The relaxation $\tau_{\rm col} =1/[\langle \sigma v\rangle n(t)]$ is estimated by a thermal average of the elastic $\pi - \pi $
cross section $\sigma \approx 23$ mb \cite{Welke}, determined by scattering phase shift data.
The thermal average  $\langle \sigma v\rangle \sim 7 - 10$ mb is nearly not depending on time, so that the collision time 
drops down proportional to $1/t$ in the first stage, but proportional to $1/t^3$ in the later stage which induces the freeze-out which occurs at $t\approx 10$ fm/c, where this transition occurs.
The relaxation of the condensate mode $\langle a_1^+\rangle$ differs from that of the normal modes.
A slower relaxation entails that this mode freezes out while the normal part is further thermalizing.   

For a recent discussion of the chemical freeze-out in the phase diagram on the basis of a kinetic criterion, see 
Ref.~\cite{Blaschke:2017lvd} and references therein.

The experimental data are well reproduced by the fit of a pion (and kaon) distribution with a nonequilibrium chemical potential as a Lagrangian multiplier in CERN SPS \cite{Kataja:1990tp}.
The thermal freeze-out process of pions and kaons at LHC conditions is characterized by just two parameters, the 
freeze-out time $\tau_{\rm fo}$ and the transverse size $r_{\rm max}$, whereby the shape of the transverse momentum spectra is described with only one parameter, $r_{\rm max}/\tau_{\rm fo}$, because the volume at freeze-out $V=\pi \tau_{\rm fo}r_{\rm max}^2$ 
fixes the overall normalization \cite{Begun:2013nga}.
For an excellent simultaneous fit of pion, kaon and proton spectra in most central Pb+Pb collisions at $\sqrt{s}=2.76$ TeV,
including the low-momentum enhancement of pions, a non-equilibrium chemical potential of pions $\mu_\pi=134.9$ MeV is required which is very close to the neutral pion mass $m_\pi=134.98$ MeV. The other parameters are   
$T_{\rm kin} = 138$ MeV, $\tau_{\rm fo}= 7.68$ fm/c and $r_{\rm max}=11.7$ fm, according to \cite{Begun:2013nga}.
A scenario as described by the Zubarev approach to the nonequilibrium statistical operator, with a fast relaxation to an intermediate relevant operator describing a Bose-Einstein condensate of pions  and the slow relaxation to full thermodynamic equilibrium seems to be realistic with these estimates of time scales.

%%%%%%%%%%%%%%%%%%%%%%%%%%%%%%%%%%%%%%%%%%%%%%%%%%%%
\section{Conclusions}			
%%%%%%%%%%%%%%%%%%%%%%%%%%%%%%%%%%%%%%%%%%%%%%%%%%%%

There are different models to describe the low momentum enhancement of pions observed in HIC at 
SPS, RHIC and LHC energies.
We discuss this effect as a signature of a quantum condensate of the high-density pion gas.
As an origin for the high phase space density of pions one may think of an initial state in the form of a color glass condensate state (gluon saturation) which gets converted to a pion gas by particle number conserving process as described, e.g.,  
\cite{Nazarova:2019dif,Harrison:2019prs}. 

After the hadronization time a hadron gas is formed, which under 
LHC conditions mainly consists of pions.  The time evolution of the fireball  is governed by 
particle-conserving binary collisions, processes which change the pion number are weak and 
influence only the long-time evolution. 
In the oversaturated pion gas the cross sections of pion rescattering processes are relatively large. 
As a consequence, the pion distribution function quickly relaxes to 
local thermodynamic equilibrium (here denoted as relevant distribution) which slowly evolves to full equilibrium.

The expansion of the fireball produced in HIC changes not only the parameter of the local thermodynamic equilibrium
but influences also the relaxation time, and freeze-out happens if the relaxation rate becomes smaller than the expansion rate.
A general description of this nonequilibrium process is given here within the Zubarev method of the
nonequilibrium statistical operator.  
As discussed in the literature, it appears that one can capture the essence of the effect with a fixed point dynamics.
The relevant statistical operator may be considered as a transient distribution proposed also recently from other approaches \cite{Erne:2018gmz,Mazeliauskas:2018yef}.
Here we formulate this behavior using the Zubarev concept of a relevant statistical operator.
The system quickly relaxes to a relevant distribution (pre-equilibrium state) which evolves slowly to equilibrium,
but is frozen out at the freeze-out time. This relevant distribution at $t_{\rm fo}$ 
describes the composition to be observed in the experiments.

We show different possibilities to introduce a relevant statistical operator to derive the corresponding equations of
evolution of the state. Treating the binary collisions in relaxation time approximation,  
a quantum condensate may appear in the relevant statistical operator. After freeze-out where this relevant distribution
stops to evolve further, the nonequilibrium evolution of the pion system is described by kinetic equations with initial condition
at freeze-out time for the distribution function which is approximated by the relevant statistical operator at $t_{\rm fo}$.
To obtain a optimum description already in lowest order perturbation theory (Markov approximation), the relevant 
statistical operator should contain already all relevant correlations, in particular the formation of quantum condensates.

The method of the Zubarev NSO as presented here for the application to the pion production in heavy-ion collision experiments considers not only a systematic description of the collision processes but also a simultaneous treatment of the hydrodynamical evolution as well as the evolution of the condensate. 

%%%%%%%%%%%%%%%%
\section*{Acknowledgements}
This work was supported by the MEPhI Academic Excellence Program under contract number  02.a03.21.0005.
%the Deutsche Forschungsgemeinschaft (DFG) under 
%contract BU 2406/1-1 and by 
%the Polish National Science Centre within the ``Maestro'' programme under 
%contract UMO-2011/02/A/ST2/00306.

%%%%%%%%%%%%%%%%%%%%%%%%%%%%%%%%%%%%%%%%%%
\end{document}